\documentclass{article}
\usepackage[utf8]{inputenc}
\usepackage{booktabs}
\usepackage{graphicx}
\usepackage[margin=1in]{geometry}
\usepackage{amsmath}
\usepackage{subcaption}
\usepackage{makecell}
\usepackage{multirow}
\usepackage{authblk}
\usepackage{tabularx}
\usepackage{float}
\usepackage{xcolor}
\usepackage[colorlinks=true, linkcolor=black, citecolor=black, urlcolor=blue]{hyperref}
\usepackage[sorting=none,style=numeric-comp]{biblatex}
\AtBeginDocument{}
\title{Comparative Evaluation of Carotid Artery Hemodynamics: Patient-Specific CFD Simulations vs. 4D flow MRI}

\author[1]{Ali Mokhtari\thanks{ali.mokhtari@unibe.ch}}
\author[1]{Merel Meulendijks}
\author[1]{Ariel Bergmann}
\author[2]{Christoph Strecker}
\author[2]{Jonathan Andrae}
\author[2]{Andreas Harloff}
\author[1]{Dominik Obrist}

\affil[1]{ARTORG Center for Biomedical Engineering Research, University of Bern, Bern, Switzerland}
\affil[2]{Department of Neurology and Neurophysiology, Faculty of Medicine, Medical Center, University of Freiburg, Freiburg, Germany}

\begin{document}
\date{}
\maketitle

\begin{abstract}
Wall shear stress (WSS) is implicated in carotid atherosclerosis, and 4D flow MRI offers a non-invasive route to its estimation, but the agreement of MRI-derived plaque-surface WSS with computational fluid dynamics (CFD) has not been quantified in a large cohort. We compared peak-systolic velocity and WSS from 4D flow MRI and patient-specific CFD in 240 carotid arteries of 120 patients, with CFD geometries and boundary conditions derived from MRI. Velocity and circumferential WSS were compared at three standardized cross-sections, and plaque-surface WSS in 63 of 149 stenosed arteries with a reliably segmented plaque. Agreement was assessed with Bland-Altman analysis, Spearman correlation with stenosis degree, and exploratory multivariable regression of the plaque-surface discrepancy on hemodynamic and geometric variables.

MRI reproduced the broad flow features but underestimated peak velocity, increasingly toward the distal internal carotid artery. WSS disagreement was location dependent: mean circumferential WSS showed little average bias in the common carotid artery, whereas MRI missed the high WSS at the bifurcation apex and underestimated plaque-surface mean and maximum WSS by $40.9 \pm 28.8\%$ and $44.5 \pm 26.8\%$; the difference in Pa increased with stenosis degree ($\rho = 0.47$ and $0.50$). In exploratory regression, the CFD-derived pressure drop across the plaque was the only significant correlate of the mean WSS discrepancy, whereas stenosis degree and geometric variables were associated with the maximum WSS discrepancy.

Under the present protocol, MRI-derived plaque-surface WSS showed magnitude- and location-dependent disagreement with CFD, greater in more severely stenosed arteries and, in exploratory analyses, associated with factors beyond stenosis degree.

\end{abstract}

\section{Introduction}

Stroke is a leading cause of death and long-term disability worldwide, and atherosclerotic carotid artery disease is a major cause of ischemic stroke~\cite{update2017heart, furie2011guidelines}. The progression of atherosclerosis is influenced not only by systemic risk factors but also by local hemodynamic forces acting on the arterial wall~\cite{malek1999hemodynamic}: low wall shear stress (WSS) and flow recirculation have been implicated in the initiation and progression of atherosclerotic lesions at the carotid bifurcation, where the common carotid artery (CCA) divides into the internal (ICA) and external (ECA) carotid arteries~\cite{ku1985pulsatile, gallo2012helical}, and elevated WSS on the plaque surface has been associated with complicated, rupture-prone plaques~\cite{Strecker2025}.

Conventional vascular imaging characterizes vessel anatomy and stenosis severity but provides limited information on local blood flow~\cite{grant2003carotid, koelemay2004systematic, debrey2008diagnostic}. Four-dimensional flow magnetic resonance imaging (4D flow MRI) captures time-resolved, three-dimensional velocity fields throughout the cardiac cycle~\cite{strater20184d, markl20124d, rothenberger2022modeling}, allows quantification of complex flow patterns~\cite{rizk20214d, itatani2022hemodynamic, markl2003time}, and permits the derivation of secondary parameters such as WSS~\cite{dyverfeldt20154d, waahlin20224d}. Its feasibility in the carotid bifurcation was demonstrated early~\cite{harloff20093d}. Computational fluid dynamics (CFD), by solving the Navier-Stokes equations in patient-specific geometries, provides velocity and WSS fields at a spatial resolution far beyond that of MRI~\cite{morbiducci2011mechanistic, ong2020computational}.

Comparisons of the two modalities have shown good qualitative agreement of velocity distributions and flow patterns in large arteries~\cite{ngo2019four, szajer2018comparison, wu2022hemodynamic, harloff20093d, mokhtari2025comparison}. Owing to its limited spatial and temporal resolution, however, 4D flow MRI underestimates peak velocities and secondary flow structures in geometrically complex regions~\cite{cibis2016effect, cherry2022impact, rothenberger2022modeling, ngo2019four}, and it systematically underestimates WSS, which depends on the near-wall velocity gradient, while broader spatial WSS patterns are largely preserved~\cite{potters20144d, szajer2018comparison, cibis2016effect, perinajova2021assessment}. The size of these discrepancies depends on voxel size, velocity encoding and the WSS estimator on the MRI side~\cite{rothenberger2022modeling, el2023optimization, markl20124d, stankovic20144d}, and on boundary conditions and modeling assumptions on the CFD side~\cite{campbell2012effect, steinman2005flow, marzo2011computational, morbiducci2011importance, vignon2010outflow, wang2025silico}.

Existing CFD-MRI comparisons have been limited to small cohorts (typically $n \leq 20$ subjects) and, to our knowledge, few studies have quantified the discrepancy in WSS at the plaque-lumen interface, the site most relevant to plaque vulnerability, in a large clinical cohort. The factors that drive inter-artery variability of this discrepancy are also poorly characterized. We therefore aimed to (1) quantify the discrepancies in peak-systolic velocity and WSS between patient-specific CFD and 4D flow MRI across 240 carotid arteries, with particular focus on plaque-surface WSS, and (2) explore which hemodynamic and geometric factors are associated with the plaque-surface WSS discrepancy.

\section{Methods}
\subsection{Study Population}
The study cohort was drawn from a previously published prospective dataset of patients aged 50 years or older with arterial hypertension and at least one carotid plaque (wall thickness $\geq 1.5$\,mm, $<50\%$ stenosis by clinical criteria at recruitment)~\cite{strecker2020carotid}. In total, 120 patients with 240 carotid arteries were included in this study. The cohort comprised 86 males and 34 females. Stenosis degree was quantified for every artery from the segmented lumen geometry (Section~\ref{sec:stenosis}); 149 arteries were classified as stenosed, with a mean stenosis degree of $34.15 \pm 17.11\%$. Because this geometry-based measure differs from the diameter criterion used at recruitment, individual values above 50\% occur. Each carotid artery was treated as the unit of analysis.

\subsection{MRI Measurements}
Four-dimensional flow MRI data were acquired on a 3\,T scanner (Prisma, Siemens Healthineers, Erlangen, Germany) with an 8-channel surface coil (NORAS MRI Products, Hoechberg, Germany) using a prospectively ECG-triggered, k-t accelerated 3D phase-contrast sequence with an isotropic spatial resolution of 0.8\,mm and a temporal resolution of 52.8\,ms~\cite{strecker2021carotid,strecker2020carotid}; the full acquisition parameters are listed in Supplementary Table~S1. Preprocessing (noise filtering, eddy-current correction, velocity anti-aliasing) was performed with the CaroTo extension of MEVISFlow (Fraunhofer MEVIS, Bremen, Germany)~\cite{Wehrum2014}. The lumen was segmented with an nnU-Net~\cite{isensee2021nnu} trained on semi-manual labels annotated by an experienced neurologist (median Dice coefficient 0.88 on the test set), and the masked phase images were converted into velocity vector fields.

\subsection{Patient-Specific CFD Simulation}
Patient-specific geometries were reconstructed from the segmented MRI data and exported as stereolithography (STL) surfaces. Segmentation-induced surface irregularities were reduced using a Taubin smoothing filter while minimizing surface shrinkage and preserving the underlying vascular morphology~\cite{taubin1995smoothing}. Centerlines were extracted using the Vascular Modeling Toolkit (VMTK, \href{http://www.vmtk.org}{www.vmtk.org}) and used to define inlet and outlet planes perpendicular to the local centerline direction. Hexahedral-dominant meshes were generated using \textit{blockMesh} and \textit{snappyHexMesh} in OpenFOAM. A maximum background cell size of 0.05\,mm was applied~\cite{stroud2002numerical}, with the local cell size reduced to 0.025\,mm in the carotid bulb and stenotic regions. Five near-wall layers were added to resolve near-wall velocity gradients, with a first-layer thickness of 0.01\,mm and a layer-to-layer growth factor of 1.2. The adequacy of this meshing strategy for WSS estimation was demonstrated through a previously published mesh-independence analysis~\cite{mokhtari2025comparison}.

Inflow conditions were derived from the phase-contrast MRI measurements~\cite{bozzi2017uncertainty}. After rigid registration of the MRI and CFD geometries by the iterative closest point algorithm~\cite{zhang2021icp}, the measured velocities were interpolated in space onto the CCA inlet and in time over the cardiac cycle. To compensate for inconsistencies caused by vessel compliance, side branches and measurement uncertainty, flow rates were averaged over four cross-sections per branch, the CCA, ICA and ECA waveforms were synchronized, and the ICA and ECA flows were scaled to the CCA waveform while preserving the measured ICA:ECA ratio~\cite{hoi2010effect}. The corrected waveform was prescribed at the ICA outlet and a zero-stress condition at the ECA outlet~\cite{morbiducci2010outflow}; walls were rigid with no-slip.

The Navier-Stokes equations were solved without a turbulence model in a quasi-direct numerical simulation (q-DNS) approach with second-order accuracy in space and time~\cite{komen2014quasi}, validated against a full direct numerical simulation (DNS) in earlier work~\cite{mokhtari2025comparison}, using the PIMPLE algorithm in OpenFOAM~\cite{weller1998tensorial, holzmann2019mathematics}. Blood was Newtonian (viscosity $0.004\,\mathrm{Pa}\cdot\mathrm{s}$, density 1060\,kg/m$^3$) and the adaptive time step kept the Courant number below 0.6. Three cardiac cycles were simulated and the first two discarded~\cite{lee2008direct}. CFD WSS was computed on the wall patch as the dynamic viscosity times the wall-normal gradient of the tangential velocity in the third cycle. Peak systole was defined as the time of maximum CCA flow rate in the MRI-derived inflow waveform, and the corresponding CFD time step was used, so that both modalities refer to the same cardiac phase; all comparisons in this study are at peak systole.

\subsection{WSS Estimation from 4D flow MRI}
WSS was estimated from the MRI velocity field using the parabolic-fitting method of Petersson et al.~\cite{petersson2012assessment}, following our previous implementation~\cite{mokhtari2025comparison}. This approach belongs to the broader class of polynomial- and spline-based methods commonly used for MRI-based WSS estimation~\cite{stalder2008quantitative,potters2015volumetric,van2015characterization,cibis2014wall,sebastien2019wall}. At each point on the segmented luminal surface, an inward-pointing normal was computed. The MRI velocity field was then interpolated at distances of $h=0.8$\,mm and $2h$ along this normal, and the tangential velocity component was determined at both sampling points. Assuming a no-slip boundary condition, the velocity at the vessel wall was set to zero. A second-order polynomial was fitted through the resulting three velocity values, and the velocity gradient at the wall was obtained from its derivative. The WSS magnitude was calculated by multiplying the wall velocity gradient by the dynamic viscosity, which was set to $0.004\,\mathrm{Pa}\cdot\mathrm{s}$.

\subsection{Cross-Sectional Slices}
\label{sliceselection}
\begin{figure}[!htb]
    \centering
    \includegraphics[height=7.5cm]{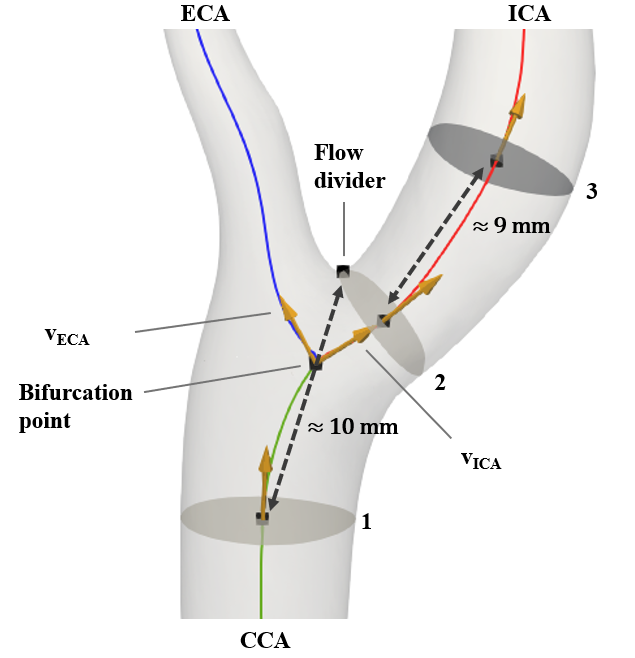}
    \caption{Definition of the three cross-sectional slices. Slice~1 lies 10\,mm upstream of the flow divider in the CCA, Slice~2 at the ICA centerline point closest to the flow divider, and Slice~3 lies 9\,mm downstream of Slice~2; $\vec{v}_{\mathrm{ICA}}$ and $\vec{v}_{\mathrm{ECA}}$ are the local branch directions. Orange arrows indicate the start and direction of the circumferential WSS extraction.}
    \label{fig:schematics}
\end{figure}

Three cross-sectional planes were defined per artery following a segmental approach~\cite{strecker2021carotid} (Figure~\ref{fig:schematics}). The bifurcation point is the transition between the ICA and ECA centerlines; the flow divider is the luminal point closest to the bifurcation point that can be reached by a positive linear combination of the branch direction vectors $\vec{v}_{\mathrm{ICA}}$ and $\vec{v}_{\mathrm{ECA}}$. Slice~1 was placed on the CCA centerline 10\,mm upstream of the flow divider along $-(\vec{v}_{\mathrm{ICA}} + \vec{v}_{\mathrm{ECA}})$, Slice~2 at the ICA centerline point closest to the flow divider, and Slice~3 was placed 9\,mm downstream of Slice~2 along the ICA centerline; each plane was orthogonal to the local centerline tangent. Velocity magnitude was sampled over the whole slice and WSS magnitude along the wall circumference of the slice plane, starting at the posterior wall and proceeding clockwise.

For qualitative comparison, velocity magnitude maps, streamlines (Supplementary Figure~S1) and circumferential WSS profiles were compared in illustrative cases: for velocity, one randomly selected artery from each stenosis tertile; for WSS, one randomly selected artery from below and one from above the median stenosis degree for each slice. For quantitative comparison, spatially averaged (mean) and maximum velocity and mean and maximum circumferential WSS were computed per slice and modality, and the per-artery differences (CFD $-$ MRI) were summarized as mean $\pm$ standard deviation (SD). Associations between the differences and stenosis degree were assessed at Slices~2 and~3, the locations prone to plaque development.

\subsection{Stenosis Quantification and Plaque Surface}
\label{sec:stenosis}
Stenosis degree and plaque surface were derived automatically from the lumen geometry using the pipeline of our companion study~\cite{mokhtari2025geometric}, which analyzes geometric and hemodynamic parameters from the same cohort and CFD simulations without comparing the modalities; the definitions are summarized here and in Supplementary Section~S6. Stenosis degree was defined as the relative area reduction $(1-A_{\min}/A_{\mathrm{ref}})\times100\%$, where $A_{\min}$ is the minimum lumen cross-sectional area at the point of maximal narrowing and $A_{\mathrm{ref}}$ is the reference lumen area distal to the stenosis. Plaque regions were identified using either roundness or luminal circumference variation, depending on the lumen morphology. For eccentric narrowing, cross-sections with roundness below 0.95 were identified as stenotic; for concentric narrowing, a relative circumference decrease of 3\% per mm marked the start of the plaque region and a corresponding increase of 3\% per mm marked its end. The luminal wall surface spanning the identified stenotic cross-sections was defined as the plaque surface.

\subsection{Plaque-Surface WSS Agreement}
Mean and maximum WSS at peak systole were evaluated on the plaque surface in both modalities. Agreement was quantified by the relative difference $(\text{WSS}_{\text{CFD}} - \text{WSS}_{\text{MRI}})/\text{WSS}_{\text{CFD}} \times 100\%$ (mean $\pm$ SD) and by Bland-Altman analysis on log$_{10}$-transformed values, in which $\log_{10}\text{WSS}_{\text{CFD}} - \log_{10}\text{WSS}_{\text{MRI}}$ is plotted against the mean of the two log values; bias and 95\% limits of agreement (mean $\pm 1.96$\,SD) were computed on the log scale and back-transformed to CFD/MRI ratios. A 95\% confidence interval (CI) for the bias was obtained by a patient-level bootstrap, in which patients were resampled with replacement so that both arteries of a patient stayed together (3000 resamples, percentile interval). Confidence intervals for the limits of agreement were not computed; the limits are reported as descriptive.

\subsection{Candidate Predictors and Multivariable Regression}
\label{sec:regression}
In the preceding analyses, stenosis degree (Section~\ref{sec:stenosis}) was used as a variable to assess its relationship with discrepancies in WSS estimation between CFD and 4D flow MRI. While stenosis degree provides a direct anatomical measure of luminal narrowing, it is unlikely to be the only factor associated with discrepancies in WSS.\\
To explore additional factors that may be associated with plaque-surface WSS discrepancies, an exploratory multivariable regression analysis was performed. A broad set of geometric and hemodynamic parameters was extracted for each artery, including:
\begin{itemize}
    \item \textbf{Geometric features:} stenosis degree, bifurcation angle, ICA angle, planarity of the ICA angle, flow divider offset, post-stenotic tortuosity, maximum curvature, and area ratios ($A_2/A_1$, $A_3/A_1$)
    \item \textbf{Hemodynamic metrics:} pressure drop across the plaque (obtained from CFD simulations), maximum velocity, ICA inflow percentage, mean helicity, mean vorticity, and turbulent kinetic energy (TKE)
\end{itemize}
These features were selected based on physiological relevance and their potential to affect local WSS distribution. All hemodynamic metrics were extracted from the CFD solution at peak systole. Definitions of the parameters are summarized in Supplementary Table~S3 (maximum velocity, mean vorticity and turbulent kinetic energy are defined in the companion study) and documented in detail in the earlier study \cite{mokhtari2025geometric}.\\
\textbf{Collinearity assessment and predictor selection} \\
Prior to model construction, all candidate predictors were screened for multicollinearity using Pearson correlation coefficients and variance inflation factors (VIFs); full results are reported in Supplementary Table~S2 and Supplementary Figure~S6 (Section~S5 of the Supplementary Material). Predictors with strong pairwise correlations (\(|r| > 0.7\)) or VIF values above 10 were considered collinear, and one variable was retained from each collinear group: mean vorticity, turbulent kinetic energy and maximum velocity were excluded. Among the strongly intercorrelated hemodynamic variables, \textit{pressure drop across the plaque} was retained based on its physiological relevance and relatively lower VIF. Between the two area ratios, only $A_2/A_1$ was retained due to its lower VIF and greater physiological relevance as a measure of ICA narrowing.\\
\textbf{Regression modeling} \\
Two separate multiple linear regression models were constructed: one for the difference in mean WSS and one for the difference in maximum WSS between CFD and 4D flow MRI on the plaque surface. The dependent variables were the signed per-artery differences (CFD minus MRI, in Pa) in plaque-surface mean and maximum WSS. The eleven selected geometric and hemodynamic parameters served as independent variables; predictors were z-standardized, so that coefficients are expressed in Pa per one-SD increase of the predictor. The models were fitted in the arteries with complete predictor data. Given the number of predictors relative to the sample size, the regression is regarded as exploratory. Model fit was summarized by \(R^2\) and adjusted \(R^2\), and two-sided \(p < 0.05\) was considered statistically significant.

\subsection{Statistical Analysis}
\label{sec:stats}
Associations between the signed CFD minus MRI differences (in m/s for velocity and Pa for WSS) and stenosis degree were quantified by Spearman's rank correlation coefficient $\rho$; given the exploratory character of these analyses, effect sizes are reported and interpreted as weak ($|\rho| < 0.4$) or moderate ($0.4 \leq |\rho| < 0.6$). Analyses were performed in Python (\texttt{scipy}, \texttt{statsmodels}).

\section{Results}

\subsection{Velocity Fields}

\begin{figure}[htbp]
    \centering
    \scalebox{0.50}{\includegraphics{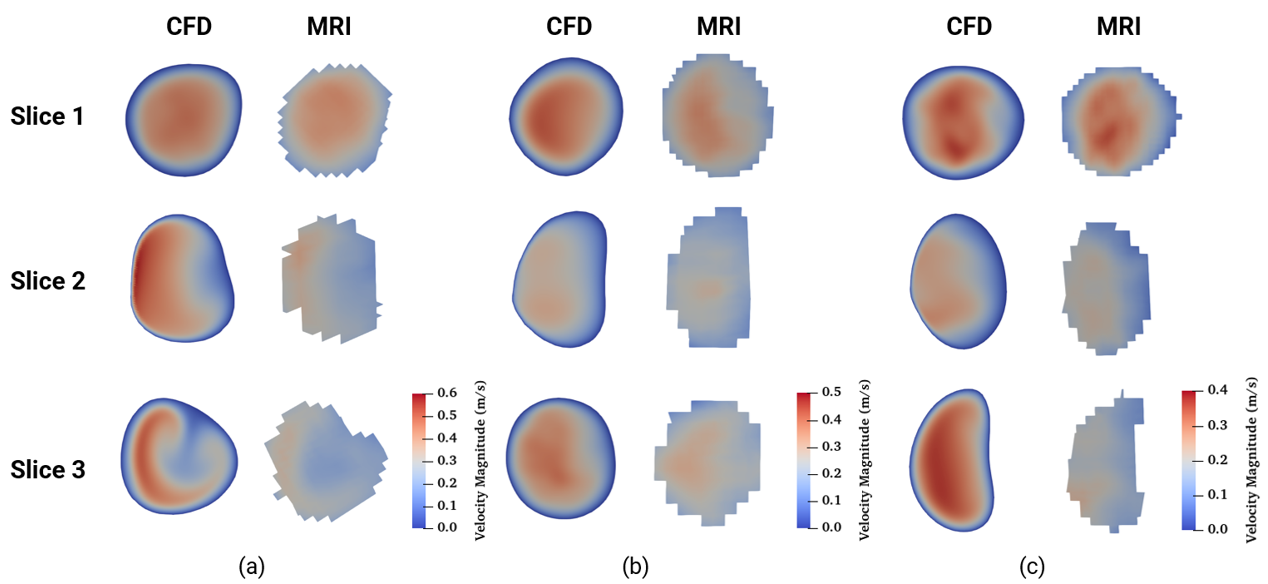}}
    \caption{Peak-systolic velocity magnitude from CFD and 4D flow MRI on the three cross-sectional slices for three illustrative arteries with increasing stenosis degree: (a)~P01 (right), 18\%; (b)~P02 (left), 30\%; (c)~P03 (left), 43\%. Color scales are matched between modalities within each case but differ between cases.}
    \label{fig:velocity fields}
\end{figure}

Figure~\ref{fig:velocity fields} shows cross-sectional velocity magnitude at the three slices for three illustrative arteries. CFD resolves sharp velocity peaks and near-wall gradients, whereas 4D flow MRI yields attenuated, spatially smoothed profiles that do not decay to zero at the lumen boundary. Peak-velocity differences increase from the CCA toward the distal ICA, where curvature, branching and narrowing produce steep gradients that MRI does not resolve, and they increase with stenosis degree (Figure~\ref{fig:velocity fields}a versus c). Streamlines (Supplementary Figure~S1) show flow separation, recirculation and helical motion in CFD near the outer bifurcation wall and in the distal ICA that are absent or attenuated in MRI.
Quantitatively (Table~\ref{tab:modality_comparison}), the mean difference in $v_\mathrm{mean}$ was $-0.030$\,m/s in the CCA, $-0.002$\,m/s in the proximal ICA and approximately zero in the distal ICA, whereas the difference in $v_\mathrm{max}$ rose from 0.084\,m/s in the CCA to 0.098 and 0.131\,m/s in the proximal and distal ICA. The velocity differences were weakly to moderately associated with stenosis degree (Supplementary Table~S4): at Slice~2, $\rho = 0.20$ for $v_\mathrm{mean}$ and $0.36$ for $v_\mathrm{max}$; at Slice~3, $0.28$ and $0.46$. The association was strongest for $v_\mathrm{max}$ at Slice~3.

\begin{table}[htbp]
\centering
\caption{Peak-systolic velocity and circumferential WSS from CFD and 4D flow MRI at the three cross-sectional slices, and their per-artery differences (CFD $-$ MRI, mean $\pm$ SD).}
\label{tab:modality_comparison}
\small
\begin{tabular}{llccc}
\toprule
\textbf{Metric} & \textbf{Slice} & \textbf{CFD} & \textbf{MRI} & \textbf{Difference} \\
\midrule
$v_\mathrm{mean}$ (m/s) & 1 (CCA) & 0.304 & 0.334 & $-0.030 \pm 0.022$ \\
 & 2 (proximal ICA) & 0.283 & 0.285 & $-0.002 \pm 0.068$ \\
 & 3 (distal ICA) & 0.289 & 0.289 & $0.000 \pm 0.073$ \\
\midrule
$v_\mathrm{max}$ (m/s) & 1 & 0.565 & 0.481 & $0.084 \pm 0.082$ \\
 & 2 & 0.523 & 0.425 & $0.098 \pm 0.114$ \\
 & 3 & 0.565 & 0.435 & $0.131 \pm 0.149$ \\
\midrule
Mean WSS (Pa) & 1 & 3.619 & 3.665 & $-0.046 \pm 3.837$ \\
 & 2 & 7.662 & 3.202 & $4.460 \pm 4.279$ \\
 & 3 & 4.384 & 3.694 & $0.690 \pm 3.367$ \\
\midrule
Max WSS (Pa) & 1 & 8.136 & 5.394 & $2.742 \pm 17.264$ \\
 & 2 & 28.009 & 5.231 & $22.778 \pm 19.011$ \\
 & 3 & 8.202 & 5.709 & $2.494 \pm 9.362$ \\
\bottomrule
\end{tabular}
\end{table}

\subsection{Circumferential WSS}

\begin{figure}[htbp]
    \centering
    \begin{subfigure}[t]{0.45\textwidth}
        \includegraphics[width=\linewidth]{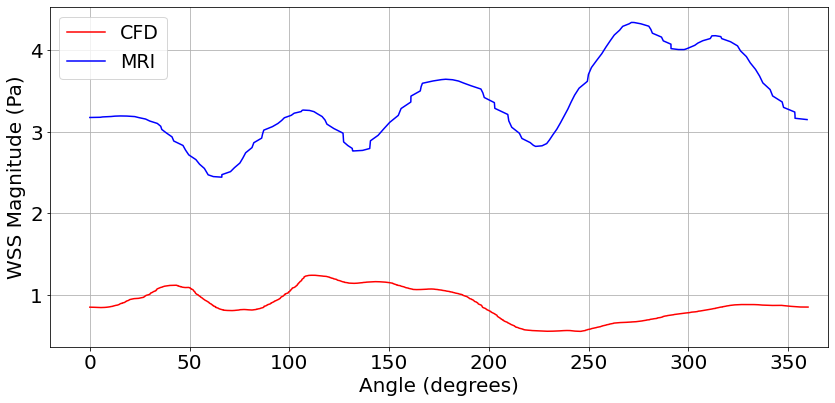}
        \caption{Slice 1: P04 (right), 27\% stenosis}
        \label{fig:P04_right_slice1}
    \end{subfigure}
    \hfill
    \begin{subfigure}[t]{0.45\textwidth}
        \includegraphics[width=\linewidth]{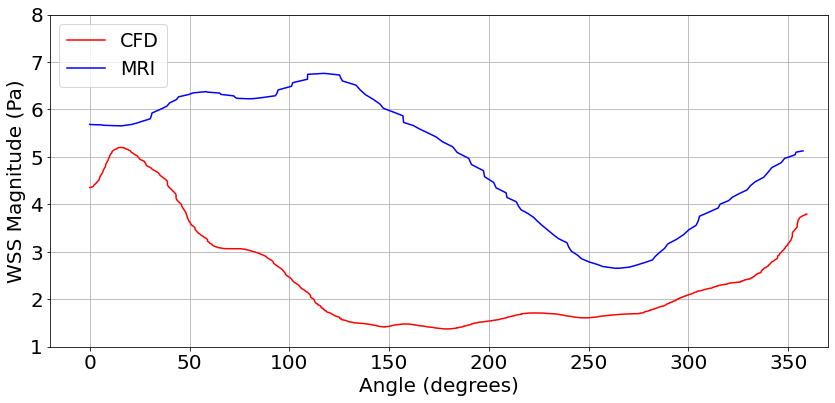}
        \caption{Slice 1: P05 (right), 50\% stenosis}
        \label{fig:P05_right_slice1}
    \end{subfigure}

    \vspace{0.2cm}
    \begin{subfigure}[t]{0.45\textwidth}
        \includegraphics[width=\linewidth]{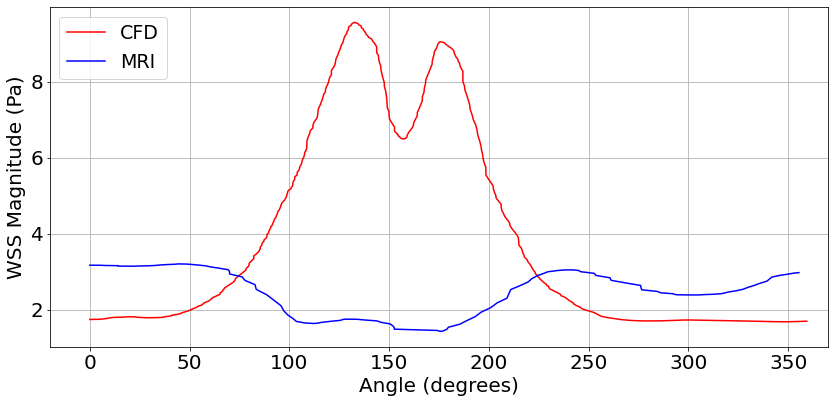}
        \caption{Slice 2: P06 (right), 29\% stenosis}
        \label{fig:P06_right_slice3}
    \end{subfigure}
    \hfill
    \begin{subfigure}[t]{0.45\textwidth}
        \includegraphics[width=\linewidth]{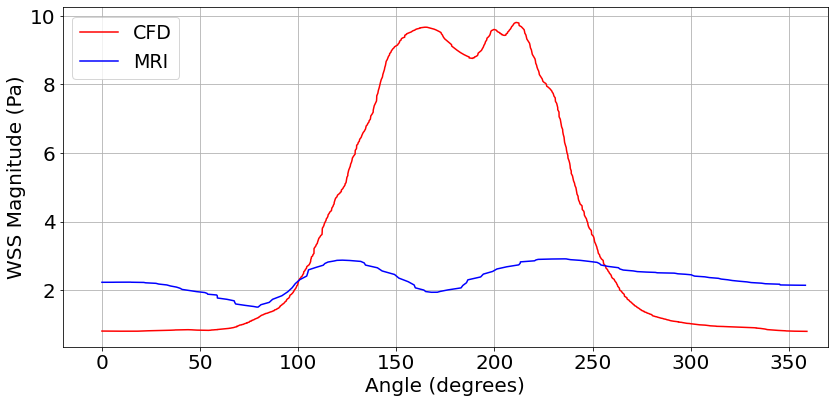}
        \caption{Slice 2: P03 (left), 43\% stenosis}
        \label{fig:P03_left_slice3}
    \end{subfigure}

    \vspace{0.2cm}
    \begin{subfigure}[t]{0.45\textwidth}
        \includegraphics[width=\linewidth]{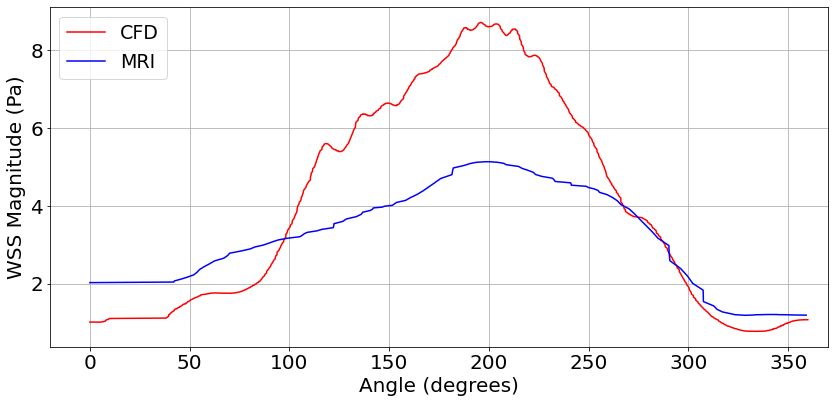}
        \caption{Slice 3: P07 (left), 14\% stenosis}
        \label{fig:P07_left_slice6}
    \end{subfigure}
    \hfill
    \begin{subfigure}[t]{0.45\textwidth}
        \includegraphics[width=\linewidth]{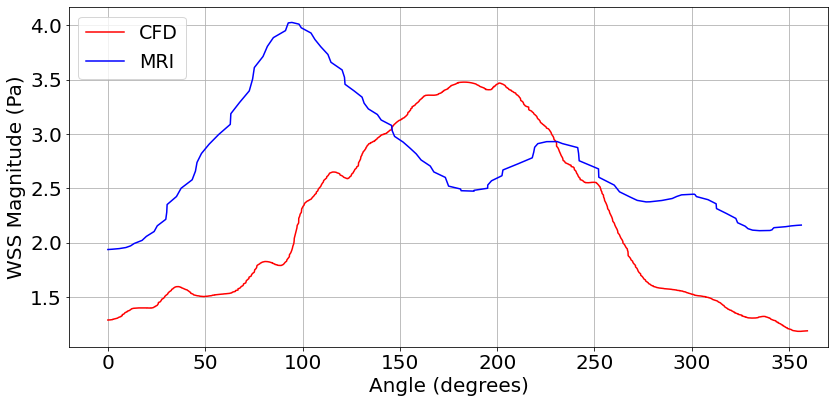}
        \caption{Slice 3: P08 (right), 31\% stenosis}
        \label{fig:P08_right_slice6}
    \end{subfigure}
    \caption{Peak-systolic WSS magnitude (Pa) along the vessel circumference at the three slices for six illustrative arteries (one below and one above the median stenosis degree per slice). The angle is measured clockwise from the posterior wall.}
    \label{fig:WSS_circumference}
\end{figure}

Figure~\ref{fig:WSS_circumference} shows circumferential WSS profiles for six illustrative arteries (14 to 50\% stenosis); full-surface WSS maps of the same arteries are given in Supplementary Figure~S2. The illustrative profiles showed three recurring patterns. In the CCA (Slice~1), the MRI-derived profiles ran above the CFD profiles along the whole circumference in these cases. Across the cohort, however, mean circumferential WSS showed little average bias at this location, whereas maximum WSS differed by about 2.7\,Pa on average; both showed considerable scatter between arteries (Table~\ref{tab:modality_comparison}). In the proximal ICA (Slice~2), CFD showed a localized WSS peak between roughly $100^{\circ}$ and $250^{\circ}$, adjacent to the bifurcation apex, that was absent from the smooth MRI profiles; on the full surface this apex peak was the most prominent and consistent difference, and it dominates the cohort statistics, where CFD exceeded MRI by about 4.5\,Pa in mean WSS and about 23\,Pa in maximum WSS at this slice. In the distal ICA (Slice~3), discrepancies varied between arteries without a consistent direction; the average differences were smaller than at Slice~2, although between-artery variability remained substantial (Table~\ref{tab:modality_comparison}, Supplementary Figure~S3). The signed CFD minus MRI WSS differences were weakly associated with stenosis degree at Slice~2 (mean WSS $\rho = 0.24$; maximum WSS $\rho = 0.16$) and at Slice~3 for mean WSS ($\rho = 0.29$), and moderately at Slice~3 for maximum WSS ($\rho = 0.45$) (Supplementary Table~S4).

\subsection{Plaque-Surface WSS}

\begin{figure}[htbp]
    \centering
    \begin{subfigure}[t]{\linewidth}
        \centering
        \includegraphics[width=0.65\linewidth]{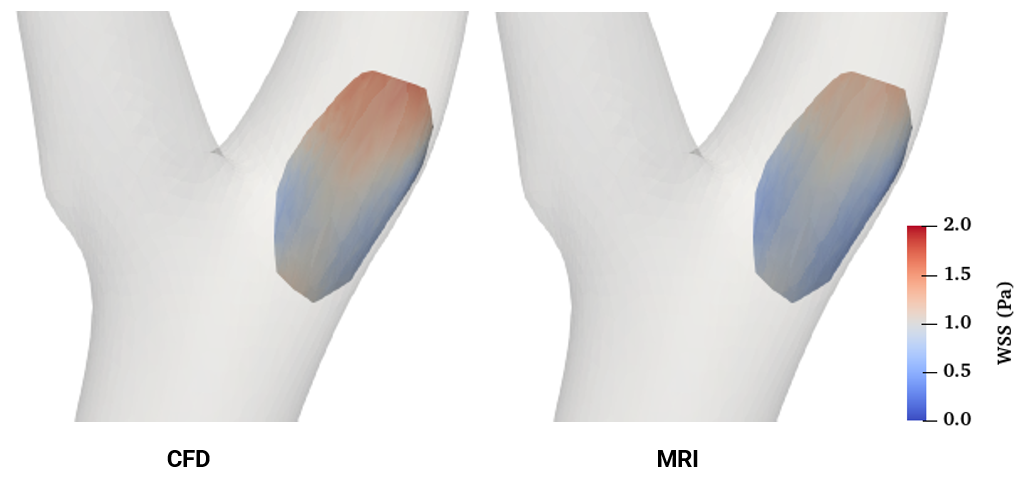}
        \caption{P03 (right), 25\% stenosis}
        \label{fig:plaque_P03_right}
    \end{subfigure}

    \vspace{0.2cm}
    \begin{subfigure}[t]{\linewidth}
        \centering
        \includegraphics[width=0.65\linewidth]{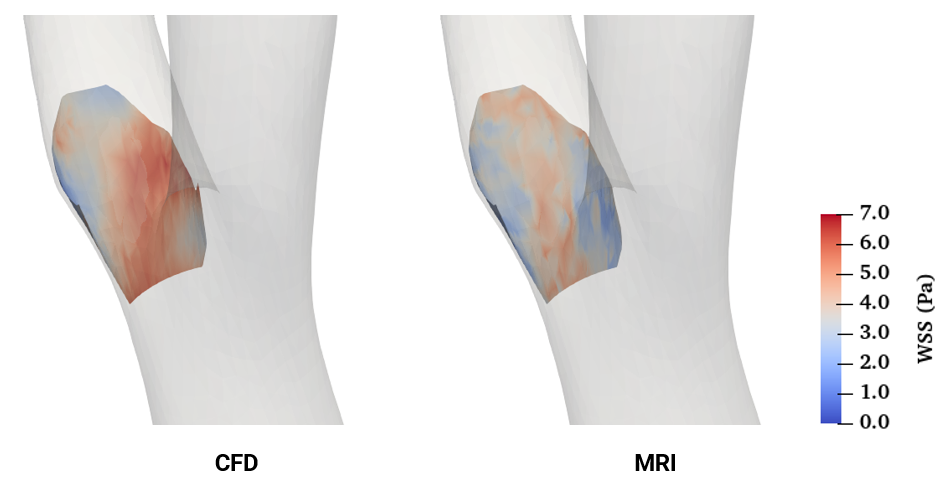}
        \caption{P09 (left), 44\% stenosis}
        \label{fig:plaque_P09_left}
    \end{subfigure}

    \vspace{0.2cm}
    \begin{subfigure}[t]{\linewidth}
        \centering
        \includegraphics[width=0.8\linewidth]{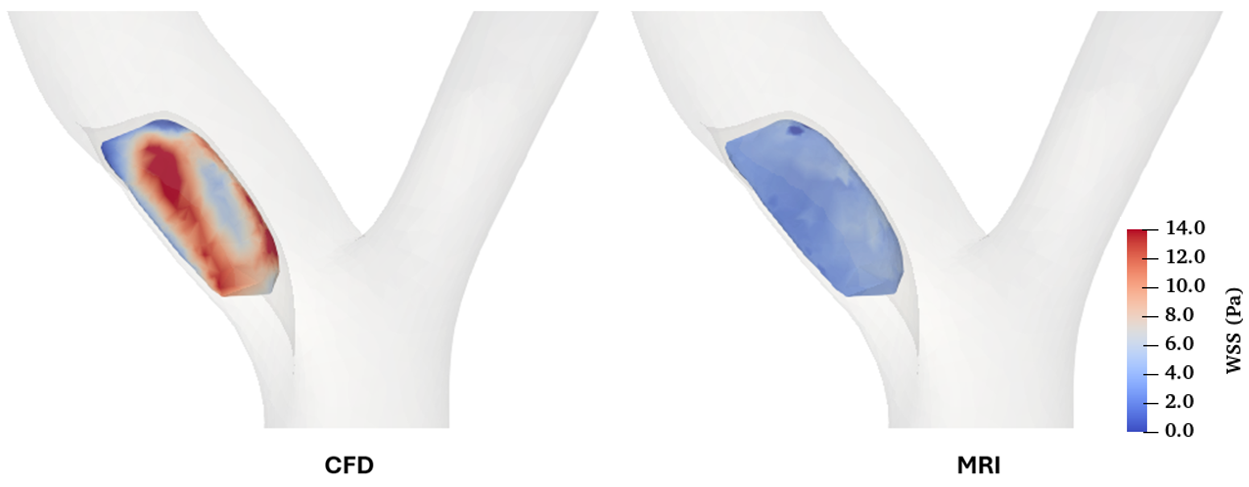}
        \caption{P10 (right), 59\% stenosis}
        \label{fig:plaque_P10_right}
    \end{subfigure}
    \caption{Peak-systolic WSS magnitude (Pa) on the segmented plaque surface from CFD (left) and 4D flow MRI (right) for three illustrative arteries. Color scales are matched between modalities within each case but differ between cases.}
    \label{fig:WSS_plaques}
\end{figure}

Figure~\ref{fig:WSS_plaques} shows plaque-surface WSS for three illustrative arteries. At 25\% stenosis (P03, right; the contralateral artery of Figure~\ref{fig:velocity fields}c) the modalities agreed in pattern and magnitude (mean and maximum differences 0.7 and 0.9\,Pa). At 44\% stenosis (P09, left) CFD showed localized high-shear regions that MRI captured only partly, with lower peaks and a patchy distribution (2.0 and 3.3\,Pa). At 59\% stenosis (P10, right) sharply defined high-WSS regions in CFD were absent or strongly attenuated in MRI (3.7 and 12.0\,Pa).

\begin{figure}[htbp]
    \centering
    \begin{subfigure}[t]{0.48\textwidth}
        \includegraphics[width=\linewidth]{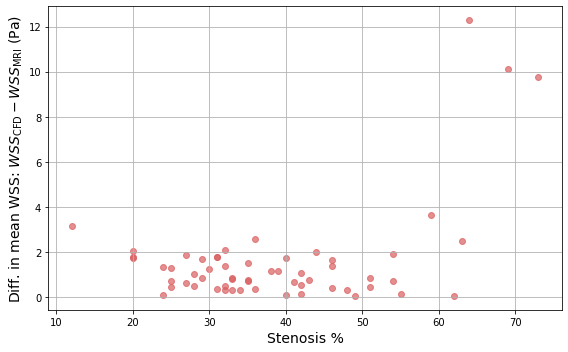}
        \caption{}
    \end{subfigure}
    \hfill
    \begin{subfigure}[t]{0.48\textwidth}
        \includegraphics[width=\linewidth]{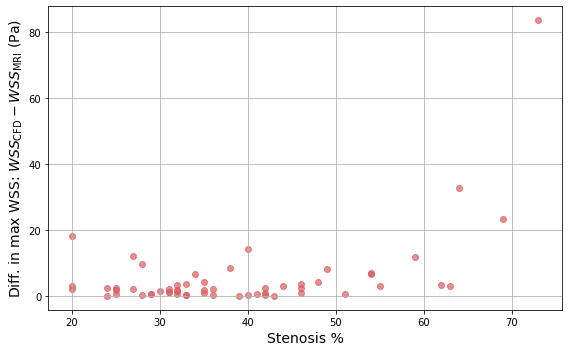}
        \caption{}
    \end{subfigure}

    \vspace{0.2cm}
\begin{subfigure}[t]{0.48\textwidth}
    \includegraphics[width=\linewidth]{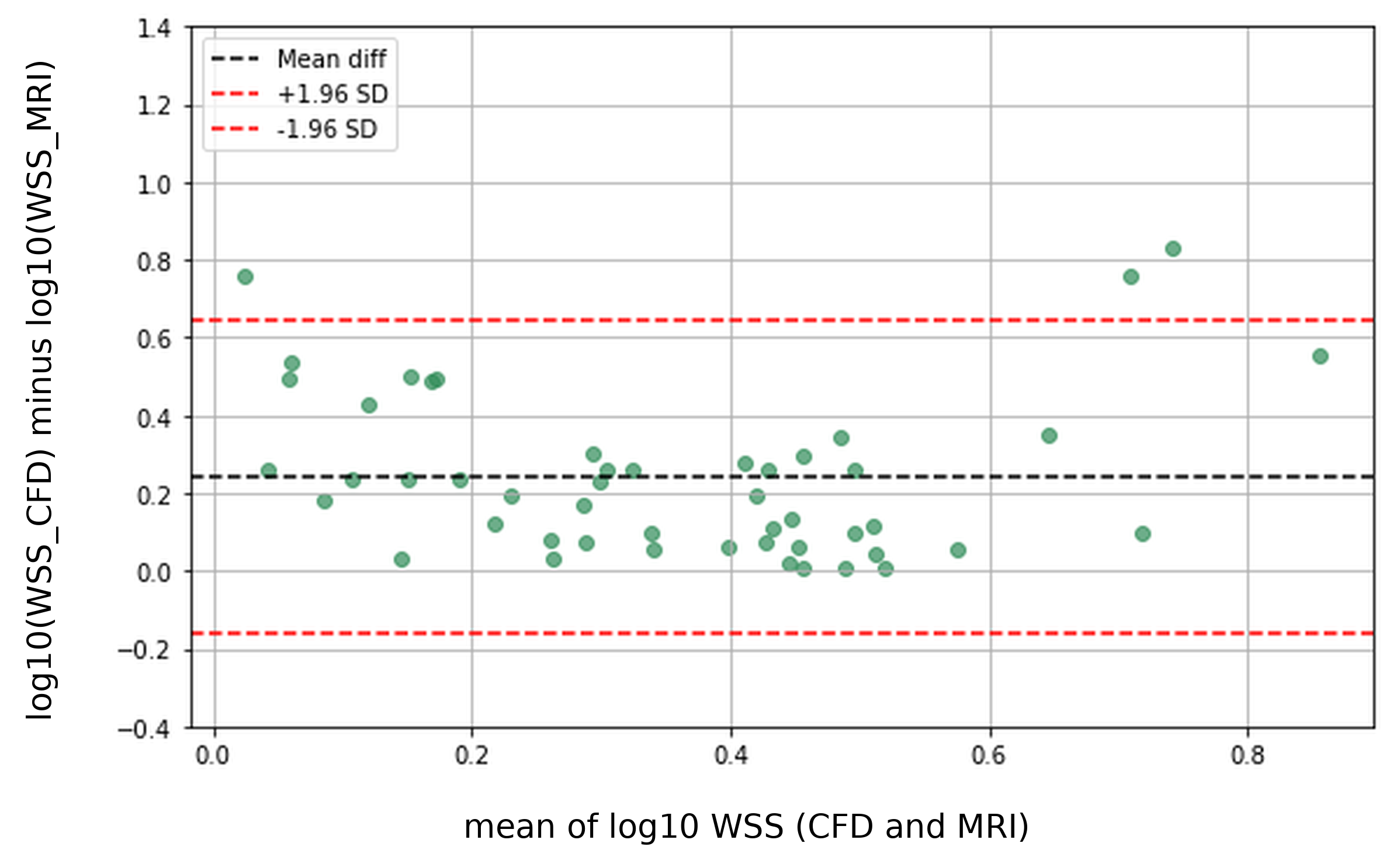}
    \caption{}
\end{subfigure}
\hfill
\begin{subfigure}[t]{0.48\textwidth}
    \includegraphics[width=\linewidth]{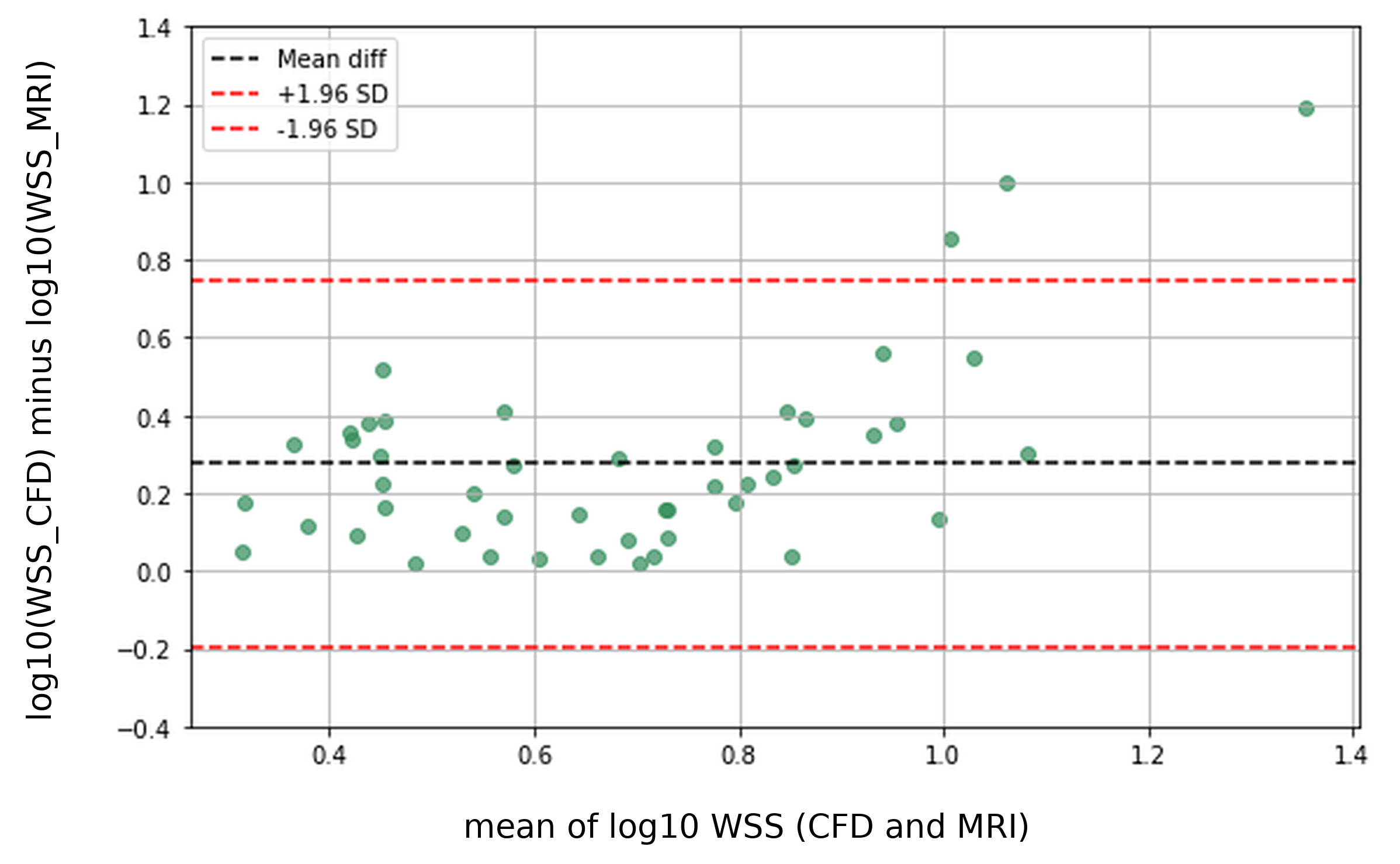}
    \caption{}
\end{subfigure}
    \caption{Agreement of plaque-surface WSS between CFD and 4D flow MRI ($n = 63$ arteries). Signed difference (CFD $-$ MRI, Pa) in (a)~mean and (b)~maximum WSS versus area-based stenosis degree (Section~\ref{sec:stenosis}). Bland-Altman plots on the $\log_{10}$ scale for (c)~mean and (d)~maximum WSS: ordinate $\log_{10}\text{WSS}_{\text{CFD}} - \log_{10}\text{WSS}_{\text{MRI}}$, abscissa the mean of the two log values; black dashed line, mean bias; red dashed lines, 95\% limits of agreement.}
    \label{fig:agreement}
\end{figure}

Across the 63 arteries, the relative difference averaged $40.9 \pm 28.8\%$ for mean WSS and $44.5 \pm 26.8\%$ for maximum WSS, i.e., MRI underestimated plaque-surface WSS by roughly 40 to 45\%. The signed difference (CFD minus MRI, in Pa) increased with stenosis degree (Figure~\ref{fig:agreement}a,b): $\rho = 0.47$ for mean WSS and $0.50$ for maximum WSS. Discrepancies tended to be larger in the few arteries above 60\% stenosis, where mean WSS differences reached about 12\,Pa and maximum WSS differences exceeded 30\,Pa. The log-scale Bland-Altman analysis (Figure~\ref{fig:agreement}c,d) showed a mean bias of $0.25$ and $0.28$ log$_{10}$ units for mean and maximum WSS, corresponding to CFD/MRI ratios of $1.8$ (95\% CI $1.42$ to $2.18$) and $1.9$ ($1.44$ to $2.54$) with descriptive 95\% limits of agreement of $0.70$ to $4.44$ and $0.64$ to $5.61$. The scatter of the log differences appeared wider at higher WSS values, so the fixed limits of agreement should be read as indicative.

\subsection{Multivariable Regression}

Multicollinearity screening excluded mean vorticity (VIF~$= 28.71$), turbulent kinetic energy (VIF~$= 13.66$) and maximum velocity (VIF~$= 10.93$) (Supplementary Section~S5). For the mean plaque-surface WSS discrepancy (Table~\ref{tab:regression}A), the model explained a substantial share of the variance ($R^2 = 0.778$, adjusted $R^2 = 0.712$, $p < 0.001$). The pressure drop across the plaque was the only variable with a statistically significant association in the exploratory multivariable model with the discrepancy ($B = 2.70$\,Pa per SD, $SE = 0.357$, $p < 0.001$); stenosis degree was not ($B = 0.127$, $p = 0.73$), and mean helicity was not statistically significant ($B = -0.539$, $p = 0.055$). For the maximum WSS discrepancy (Table~\ref{tab:regression}B), the model explained less variance ($R^2 = 0.511$, adjusted $R^2 = 0.365$, $p = 0.002$); stenosis degree ($B = 17.9$, $p = 0.001$), bifurcation angle ($B = 13.5$, $p = 0.006$), ICA inflow percentage ($B = -8.5$, $p = 0.023$) and the area ratio $A_2/A_1$ ($B = 9.1$, $p = 0.041$) were associated with it, whereas pressure drop was not.

\begin{table}[htbp]
\centering
\caption{Multiple linear regression of the plaque-surface WSS discrepancy (CFD $-$ MRI, Pa) on eleven geometric and hemodynamic variables, fitted in the arteries with complete predictor data. $B$: coefficient in Pa per one-SD increase of the predictor.}
\label{tab:regression}
\small
\begin{tabular}{lccc}
\toprule
\textbf{Predictor} & $B$ (Pa/SD) & \textbf{SE} & $p$ \\
\midrule
\multicolumn{4}{l}{\textbf{A. Mean WSS discrepancy} ($R^2 = 0.778$, adjusted $R^2 = 0.712$)} \\
Pressure drop across plaque & 2.702  & 0.357 & $<$0.001 \\
Mean helicity              & $-$0.539 & 0.272 & 0.055 \\
Bifurcation angle          & 0.169  & 0.342 & 0.623 \\
ICA angle                  & 0.036  & 0.317 & 0.909 \\
Planarity of ICA angle     & 0.245  & 0.274 & 0.377 \\
Flow divider offset        & 0.082  & 0.345 & 0.814 \\
Maximum curvature          & 0.157  & 0.283 & 0.582 \\
Stenosis degree            & 0.127  & 0.368 & 0.731 \\
ICA inflow percentage      & $-$0.415 & 0.265 & 0.127 \\
Post-stenotic tortuosity   & $-$0.352 & 0.279 & 0.215 \\
Area ratio $A_2/A_1$       & 0.223  & 0.320 & 0.490 \\
\midrule
\multicolumn{4}{l}{\textbf{B. Maximum WSS discrepancy} ($R^2 = 0.511$, adjusted $R^2 = 0.365$)} \\
Stenosis degree            & 17.868  & 4.963 & 0.001 \\
Bifurcation angle          & 13.477  & 4.616 & 0.006 \\
ICA inflow percentage      & $-$8.518  & 3.584 & 0.023 \\
Area ratio $A_2/A_1$       & 9.142   & 4.319 & 0.041 \\
ICA angle                  & $-$4.207  & 4.278 & 0.332 \\
Planarity of ICA angle     & $-$3.029  & 3.704 & 0.419 \\
Flow divider offset        & $-$1.859  & 4.660 & 0.692 \\
Maximum curvature          & $-$4.978  & 3.818 & 0.200 \\
Pressure drop across plaque & 3.637  & 4.816 & 0.455 \\
Mean helicity              & $-$4.751  & 3.668 & 0.203 \\
Post-stenotic tortuosity   & $-$3.857  & 3.766 & 0.312 \\
\bottomrule
\end{tabular}
\end{table}

\clearpage
\section{Discussion}

In 240 patient-specific carotid arteries, 4D flow MRI reproduced the broad flow characteristics seen in CFD but showed location- and magnitude-dependent disagreement for peak-systolic velocity and WSS. Plaque-surface WSS disagreement increased with stenosis degree, and exploratory regression indicated that hemodynamic and geometric factors beyond stenosis degree were associated with it. Because CFD served as a high-resolution computational reference whose geometry and boundary conditions were themselves derived from the same MRI examination, these discrepancies reflect the assumptions on both sides of the comparison.

\subsection{Velocity}
The underestimation of peak velocity by MRI grew from the CCA toward the distal ICA and with stenosis degree, in line with the loss of steep spatial gradients in millimeter-scale voxels~\cite{cibis2016effect, rothenberger2022modeling}. Peak ICA velocity is central to duplex-based stenosis grading~\cite{grant2003carotid}, so a stenosis-dependent rather than fixed underestimation matters most in the arteries where accurate assessment is most needed. Compared with earlier cohorts of at most 20 subjects~\cite{szajer2018comparison, ngo2019four}, the present cohort allows this relationship to be examined in a much larger sample.

\subsection{Circumferential WSS and the No-Slip Artifact}
Two distinct error modes were observed. At the bifurcation apex (Slice~2), CFD resolved a localized WSS peak that MRI missed entirely, reflecting unresolved near-wall gradients generated by branching and flow separation~\cite{potters20144d, cibis2016effect}. In the CCA (Slice~1), by contrast, mean circumferential WSS showed little average bias across the cohort, although maximum WSS still differed on average and the individual differences scattered widely in both directions, and in the illustrative profiles MRI-derived WSS ran above CFD along the whole circumference. A plausible explanation for such local overestimation is the no-slip condition imposed in the parabolic fit: with the wall not explicitly resolved, forcing the fit through zero at a coarsely sampled boundary steepens the near-wall gradient where the true gradient is shallow~\cite{gao2025postprocessing}. This mechanism is illustrated in one artery (Supplementary Section~S4) but was not tested systematically; applying the MRI estimator to a downsampled CFD field would separate resolution from post-processing effects and is left to future work. The coexistence of underestimation in complex flow regions and overestimation in some laminar segments means that the sign and size of the MRI WSS error depend on local geometry and flow regime, which suggests that a single global correction may be inadequate and underlines that MRI-derived WSS is a model-dependent estimate sensitive to segmentation, wall identification and interpolation.

\subsection{Plaque-Surface WSS}
The underestimation at the plaque surface (about 40 to 45\%) exceeded that reported at mid-vessel locations in previous smaller studies~\cite{szajer2018comparison, cibis2016effect, ngo2019four}. Plaques create abrupt geometric changes with steep gradients, separation and recirculation that are beyond the resolving capacity of the voxels, and the present cohort spans a wider stenosis range than previous validation studies. Resolution-dependent WSS errors of similar character have been reported in the aorta~\cite{perinajova2021assessment}. Part of the maximum WSS discrepancy is expected on statistical grounds alone: the maximum of a densely sampled CFD surface field is an extreme-value statistic that can exceed the maximum of a coarsely sampled, smoothed field even without measurement error. CFD-MRI disagreement was greatest at the high-shear values that have been associated with complicated plaques~\cite{Strecker2025}. This is potentially clinically relevant because it may limit comparison of absolute plaque-surface WSS values across modalities; classification of plaques or clinical outcomes was not evaluated here.

\subsection{Correlates of the Discrepancy}
In the exploratory regression, the pressure drop across the plaque was the only variable with a statistically significant association in the exploratory multivariable model with the mean WSS discrepancy, whereas stenosis degree was not. Pressure drop integrates the hemodynamic consequences of plaque length, shape, eccentricity and inflow, so two plaques with equal stenosis degree can produce different near-wall gradients that MRI may or may not resolve. For the maximum WSS discrepancy, stenosis degree, bifurcation angle, ICA inflow percentage and the area ratio $A_2/A_1$ were associated, and less variance was explained, consistent with the sensitivity of peak WSS to local features not captured by global descriptors. Three caveats apply. First, with eleven predictors in a modest number of arteries the coefficients are imprecise and the model is exploratory. Second, the pressure drop was obtained from CFD, and both it and the CFD WSS are functions of the same simulated flow field, so the two quantities are physically and computationally coupled; pressure drop is therefore an exploratory correlate, not a validated predictor. Third, whether an MRI-derived pressure drop, obtainable from 4D flow velocity fields~\cite{dyverfeldt20154d}, carries the same information was not tested. Only if these associations are confirmed in independent data with MRI-derived pressure estimates could a pressure-informed adjustment of MRI WSS, or hybrid MRI-CFD and physics-informed learning approaches~\cite{ferdian2022wssnet, zhang2022wall} trained on paired datasets, be explored; the present study does not establish such a correction.

\section{Limitations}
The cohort consisted predominantly of mildly to moderately stenosed arteries from a single-center hypertensive plaque population, which limits generalizability to high-grade stenosis and other populations. Both arteries of a patient were analyzed and are not fully independent; this was accounted for in the Bland-Altman confidence intervals through a patient-level bootstrap, whereas the correlation and regression analyses treat arteries as independent observations. All comparisons were made at peak systole; time-averaged WSS, oscillatory shear and waveform features were not assessed. Although the 0.8\,mm isotropic resolution is relatively fine for clinical 4D flow MRI, it does not resolve near-wall gradients, and the parabolic-fit estimator with imposed no-slip may introduce artifacts where the wall is imprecisely segmented. CFD assumed rigid walls and Newtonian rheology and served as a computational reference rather than a validated gold standard; moreover, its geometry and boundary conditions were derived from the same MRI examination, so the two modalities share segmentation and flow information and are not fully independent. Plaque-surface WSS was evaluable in only 63 of 149 stenosed arteries and the regression used the subset with complete predictor data, so with eleven predictors the regression is exploratory and may be affected by selection if excluded arteries differ from included ones. All data came from one scanner and protocol; residual registration and boundary condition mismatches may have contributed to the reported differences.

\section{Conclusion}
In a large patient-specific carotid cohort, 4D flow MRI reproduced broad flow characteristics but showed location- and magnitude-dependent disagreement with CFD for peak-systolic WSS: little average bias in mean WSS in the CCA, a missed high-WSS peak at the bifurcation apex, and average underestimation of plaque-surface WSS by roughly 40 to 45\%, with the signed differences in Pa increasing with stenosis degree. Exploratory regression indicated that hemodynamic and geometric factors beyond stenosis degree, in particular the CFD-derived pressure drop across the plaque for the mean WSS discrepancy, were associated with this disagreement. Under the present acquisition and post-processing protocol, absolute plaque-surface WSS values from 4D flow MRI should be interpreted cautiously and should not be considered interchangeable with CFD-derived values; whether pressure-informed adjustments or hybrid MRI-CFD approaches can reduce the disagreement remains to be tested.

\section*{Funding}
This work was supported by the Swiss National Science Foundation (SNSF, grant no.\ 205321L\_197189), the Deutsche Forschungsgemeinschaft (DFG, grant no.\ HA~5399/6-1), and the Berta-Ottenstein Programme for Advanced Clinician Scientists of the Medical Faculty of the University of Freiburg. Computing time was provided by the Swiss National Supercomputing Centre (CSCS) under project ID~s1290.

\section*{Data and code availability}
The processed datasets and the analysis scripts used in this study are available from the corresponding author upon reasonable request. The raw 4D flow MRI data cannot be made publicly available because of patient privacy restrictions.

\section*{Declarations}
\textbf{Conflict of interest.} The authors declare that they have no financial or non-financial competing interests relevant to this work.\\
\textbf{Ethical approval and consent to participate.} The imaging studies from which the carotid 4D flow MRI dataset was derived were approved by the Ethics Committee of the University of Freiburg, Germany, and written informed consent was obtained from all participants. The work was conducted in accordance with the Declaration of Helsinki. The present manuscript reports a secondary, retrospective analysis of fully anonymized data from those studies.\\
\textbf{Consent for publication.} Not applicable; the manuscript contains no identifiable data of any individual.

\section*{Author contributions}
\textbf{Ali Mokhtari}: Conceptualization, Methodology, Software, Formal analysis, Investigation, Data curation, Visualization, Writing (original draft), Writing (review and editing).\\
\textbf{Merel Meulendijks}: Conceptualization, Methodology, Software, Formal analysis, Investigation, Data curation, Visualization, Writing (original draft), Writing (review and editing).\\
\textbf{Ariel Bergmann}: Conceptualization, Methodology, Software, Data curation, Writing (review and editing).\\
\textbf{Christoph Strecker}: Conceptualization, Methodology, Investigation (patient recruitment and MRI measurements), Resources, Data curation, Writing (review and editing).\\
\textbf{Jonathan Andrae}: Conceptualization, Methodology, Investigation (patient recruitment and MRI measurements), Resources, Data curation, Writing (review and editing).\\
\textbf{Andreas Harloff}: Conceptualization, Methodology, Investigation (patient recruitment and MRI measurements), Resources, Data curation, Funding acquisition, Writing (review and editing).\\
\textbf{Dominik Obrist}: Conceptualization, Methodology, Funding acquisition, Supervision, Writing (review and editing).

\printbibliography

\end{document}

\date{}
\maketitle

\noindent
This document contains supplementary tables and figures for the main manuscript. Patient cases are referred to by the anonymized codes P01 to P10 used in the main manuscript. All velocity and WSS values refer to peak systole.

% ─────────────────────────────────────────────────────────────────────────────
\section*{S1\quad 4D flow MRI Acquisition Parameters}
% ─────────────────────────────────────────────────────────────────────────────

\begin{table}[H]
    \centering
    \caption{Imaging parameters of the 4D flow MRI acquisition (3\,T Prisma, Siemens Healthineers; 8-channel surface coil, NORAS MRI Products)~\cite{strecker2020carotid,strecker2021carotid}.}
    \label{stab:4Dflow_params}
    \begin{tabular}{l c}
        \toprule
        Parameter & Value \\
        \midrule
        Spatial resolution & 0.8 mm isotropic ($0.8 \times 0.8 \times 0.8$ mm$^3$) \\
        Temporal resolution & 52.8 ms \\
        TR/TE & 52.8/3.9 ms \\
        Flip angle & 12$^\circ$ \\
        k-t GRAPPA acceleration factor & 5 \\
        Field of view & 140 $\times$ 140 mm$^2$ \\
        Bandwidth & 460 Hz/Px \\
        VENC (in-plane) & 0.6 m/s \\
        VENC (through-plane) & 1.0 m/s \\
        \bottomrule
    \end{tabular}
\end{table}

% ─────────────────────────────────────────────────────────────────────────────
\section*{S2\quad Velocity: Streamlines}
% ─────────────────────────────────────────────────────────────────────────────

\begin{figure}[H]
    \centering
    \begin{subfigure}[t]{0.30\textwidth}
        \includegraphics[width=\linewidth]{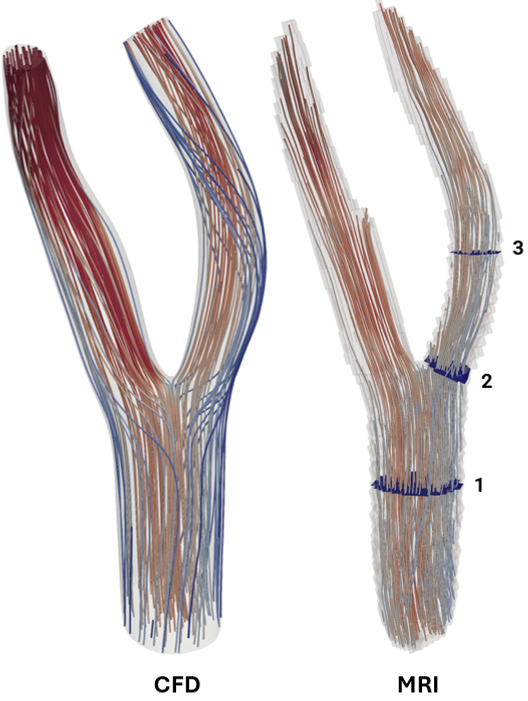}
        \caption{P01 (right), 18\% stenosis}
    \end{subfigure}
    \hfill
    \begin{subfigure}[t]{0.32\textwidth}
        \includegraphics[width=\linewidth]{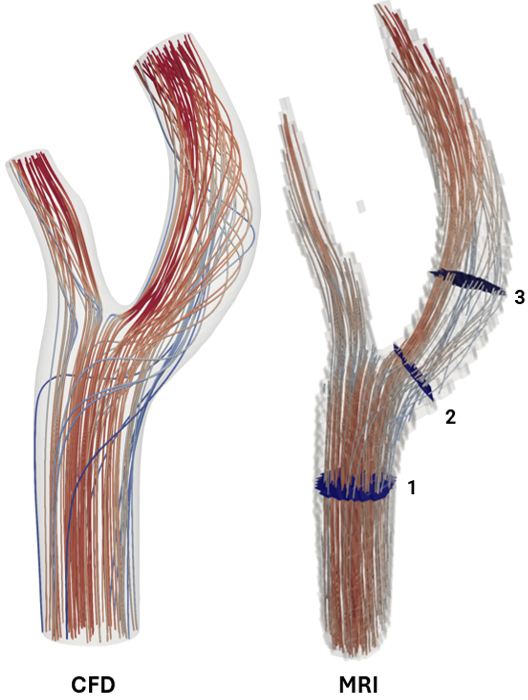}
        \caption{P02 (left), 30\% stenosis}
    \end{subfigure}
    \hfill
    \begin{subfigure}[t]{0.32\textwidth}
        \includegraphics[width=\linewidth]{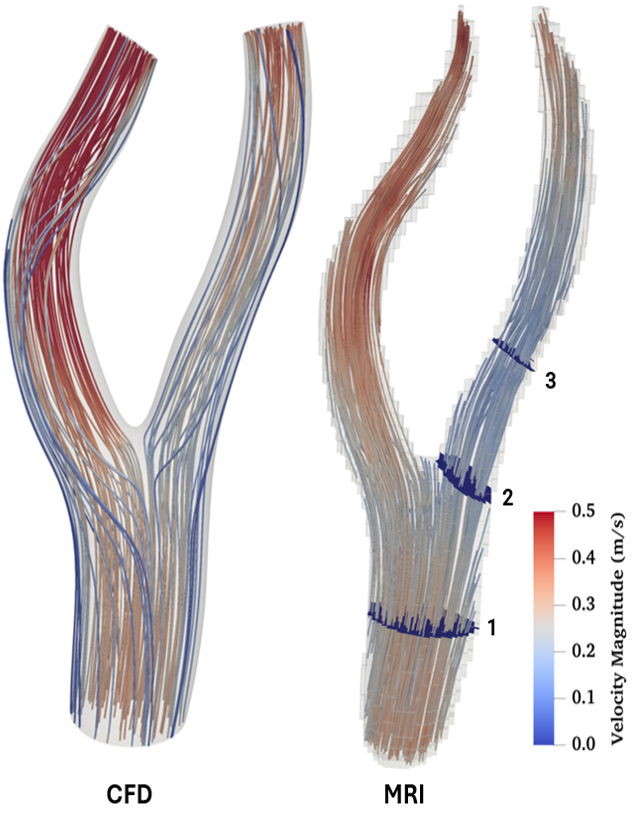}
        \caption{P03 (left), 43\% stenosis}
    \end{subfigure}
    \caption{Peak-systolic streamlines from CFD (left) and 4D flow MRI (right) for three illustrative arteries, colored by velocity magnitude. The three slice locations are indicated in the MRI plots.}
    \label{sfig:streamlines}
\end{figure}

Figure~\ref{sfig:streamlines} shows streamlines for the three illustrative arteries of Figure~2 of the main manuscript. In the CCA, streamlines are uniform and aligned in both modalities. Toward the bifurcation and in the ICA, CFD exhibits flow separation, recirculation and helical motion near the outer bifurcation wall and in the distal ICA that are absent or only partially represented in the 4D flow MRI streamlines.

\newpage
% ─────────────────────────────────────────────────────────────────────────────
\section*{S3\quad Circumferential WSS: Full-Surface Maps and Association with Stenosis Degree}
% ─────────────────────────────────────────────────────────────────────────────

\begin{figure}[H]
    \centering
    \begin{subfigure}[t]{0.28\textwidth}
        \includegraphics[width=\linewidth]{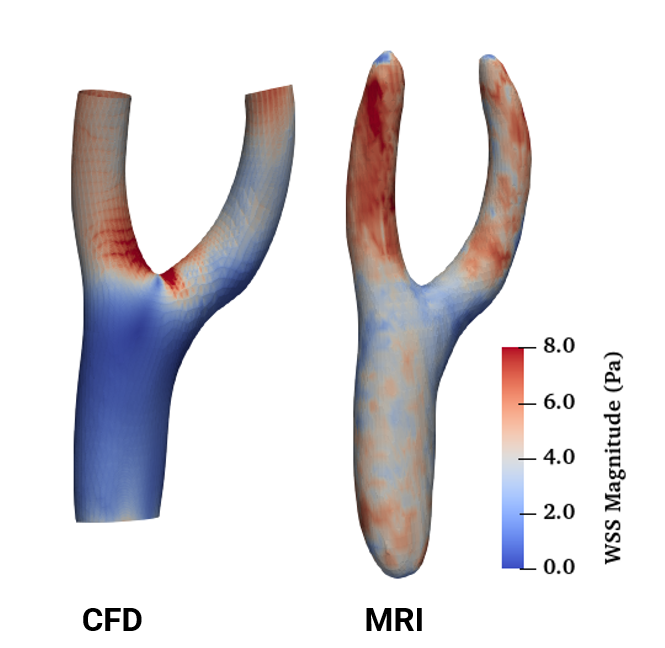}
        \caption{P04 (right)}
    \end{subfigure}
    \hfill
    \begin{subfigure}[t]{0.28\textwidth}
        \includegraphics[width=\linewidth]{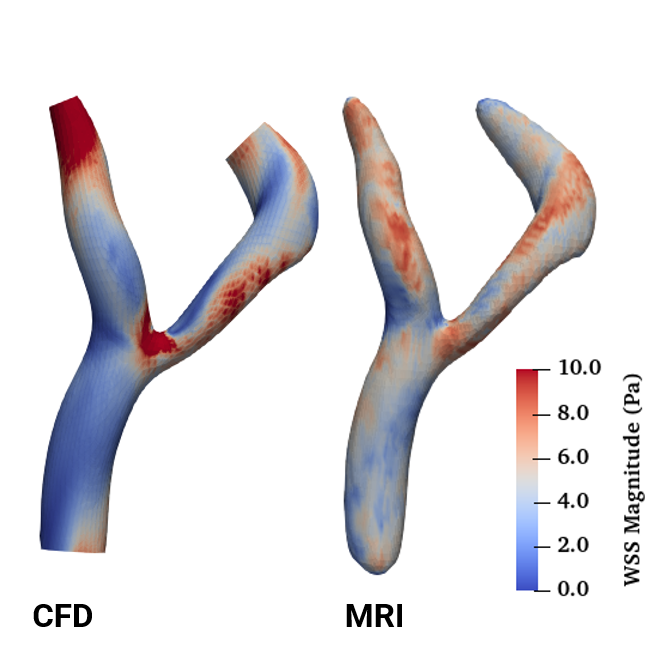}
        \caption{P05 (right)}
    \end{subfigure}
    \hfill
    \begin{subfigure}[t]{0.28\textwidth}
        \includegraphics[width=\linewidth]{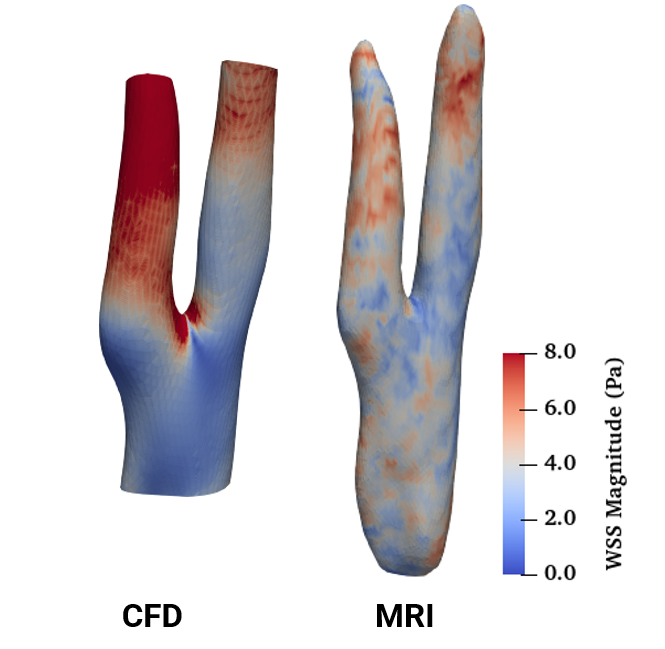}
        \caption{P06 (right)}
    \end{subfigure}

    \vspace{0.4cm}
    \begin{subfigure}[t]{0.28\textwidth}
        \includegraphics[width=\linewidth]{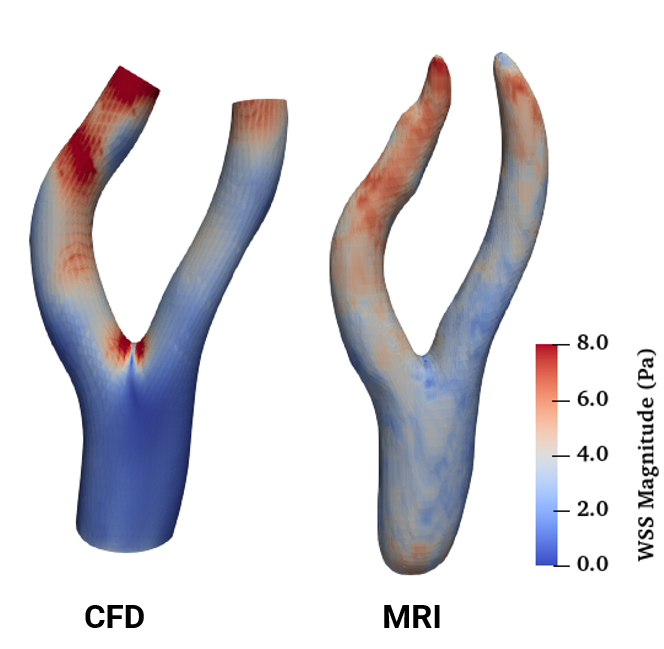}
        \caption{P03 (left)}
    \end{subfigure}
    \hfill
    \begin{subfigure}[t]{0.28\textwidth}
        \includegraphics[width=\linewidth]{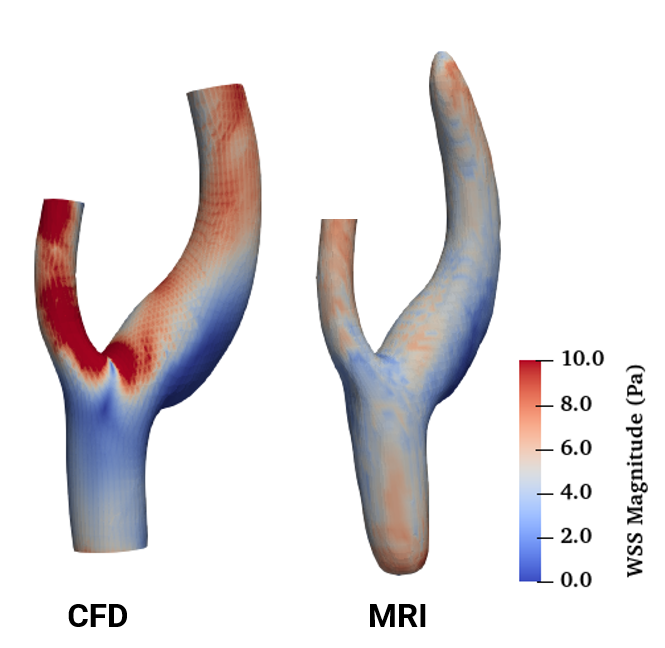}
        \caption{P07 (left)}
    \end{subfigure}
    \hfill
    \begin{subfigure}[t]{0.28\textwidth}
        \includegraphics[width=\linewidth]{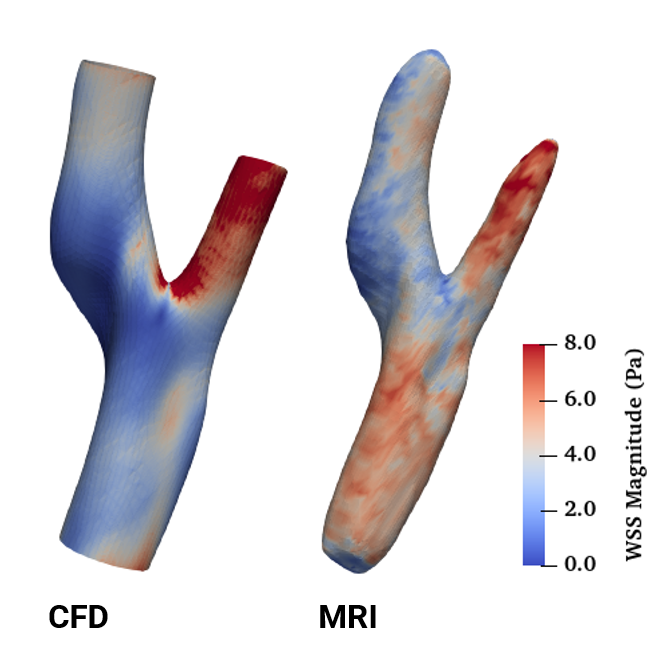}
        \caption{P08 (right)}
    \end{subfigure}
    \caption{Full-surface peak-systolic WSS magnitude (Pa) from CFD (left half of each panel) and 4D flow MRI (right half) for the six illustrative arteries of Figure~3 of the main manuscript. The color scale is shared between CFD and MRI within each panel but differs between panels.}
    \label{sfig:WSS_maps}
\end{figure}

\begin{figure}[H]
    \centering
    \begin{subfigure}[t]{0.48\textwidth}
        \includegraphics[width=\linewidth]{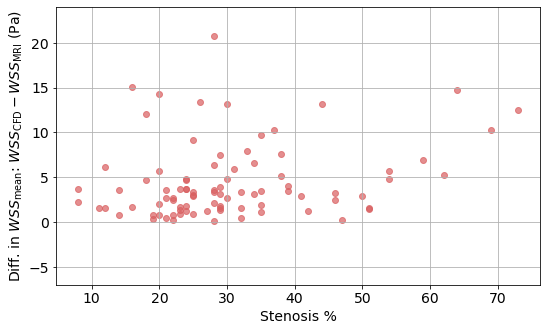}
        \caption{Mean WSS, Slice 2}
    \end{subfigure}
    \hfill
    \begin{subfigure}[t]{0.48\textwidth}
        \includegraphics[width=\linewidth]{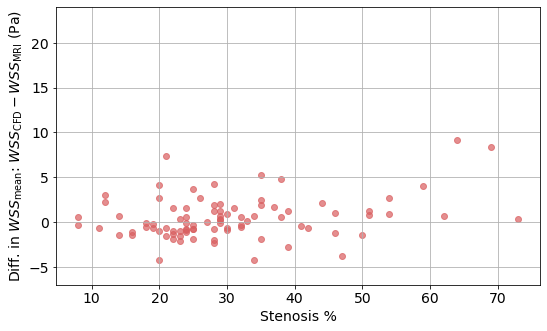}
        \caption{Mean WSS, Slice 3}
    \end{subfigure}

    \vspace{0.3cm}
    \begin{subfigure}[t]{0.48\textwidth}
        \includegraphics[width=\linewidth]{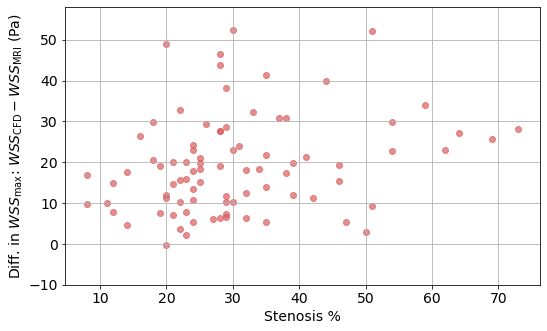}
        \caption{Maximum WSS, Slice 2}
    \end{subfigure}
    \hfill
    \begin{subfigure}[t]{0.48\textwidth}
        \includegraphics[width=\linewidth]{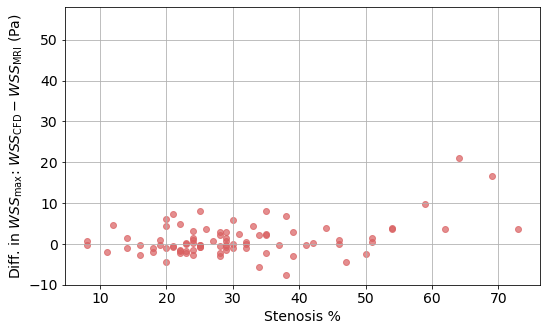}
        \caption{Maximum WSS, Slice 3}
    \end{subfigure}
    \caption{Difference (CFD $-$ MRI) in peak-systolic circumferential mean WSS (a,~b) and maximum WSS (c,~d) versus area-based stenosis degree at Slices~2 and~3.}
    \label{sfig:error_wss}
\end{figure}

Figure~\ref{sfig:WSS_maps} shows full-surface WSS maps for the six illustrative arteries of Figure~3 of the main manuscript. The elevated WSS at the bifurcation apex in CFD is absent or markedly reduced in the MRI maps. CFD fields are smooth and spatially continuous, whereas MRI maps are more heterogeneous and patchy; in the CCA, where CFD indicates low WSS, MRI tends to show somewhat elevated values in these cases, consistent with the Slice~1 profiles of Figure~3. Figure~\ref{sfig:error_wss} shows the per-artery circumferential WSS differences against stenosis degree at Slices~2 and~3.

\newpage
% ─────────────────────────────────────────────────────────────────────────────
\section*{S4\quad Impact of the No-Slip Condition on MRI-Derived WSS (Single-Case Illustration)}
% ─────────────────────────────────────────────────────────────────────────────

In several arteries, including the illustrative cases of Figure~3 of the main manuscript, Slice~1 (CCA) exhibited higher MRI-derived than CFD WSS (Section~3.2 of the main manuscript). This section illustrates, for one artery, how the no-slip condition imposed in the parabolic fit can produce this behavior; the analysis was not performed systematically across the cohort and indicates a plausible mechanism rather than an established source of error.

\begin{figure}[H]
    \centering
    \includegraphics[width=0.35\linewidth]{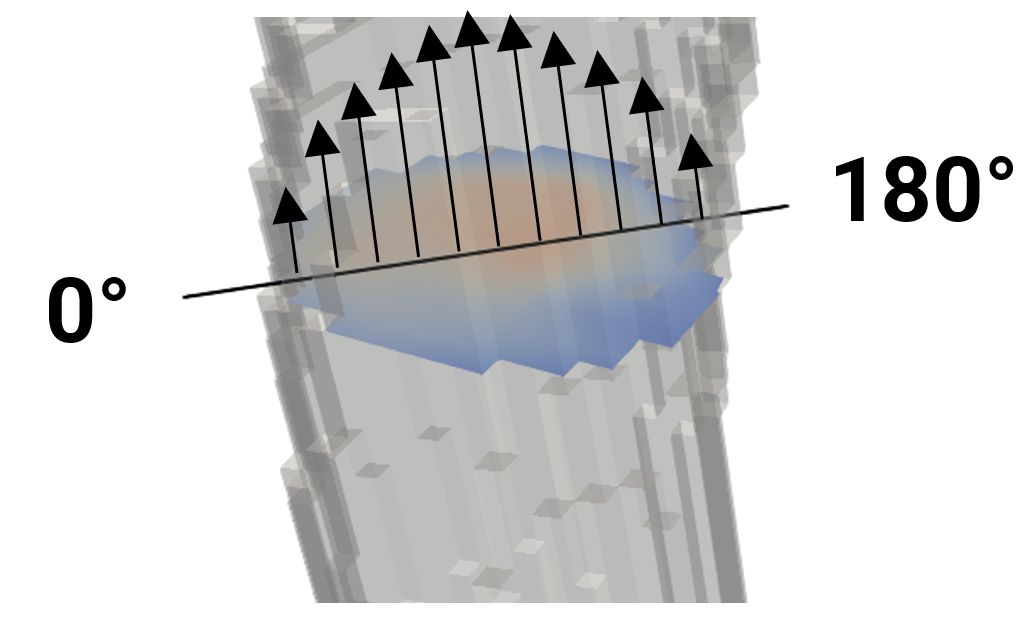}
    \caption{Extraction of the velocity profile from the vessel wall at $0^{\circ}$ across the cross-section to the opposite wall at $180^{\circ}$.}
    \label{sfig:line_velocity}
\end{figure}

\begin{figure}[H]
    \centering
    \begin{subfigure}[t]{0.47\textwidth}
        \includegraphics[height=4.5cm]{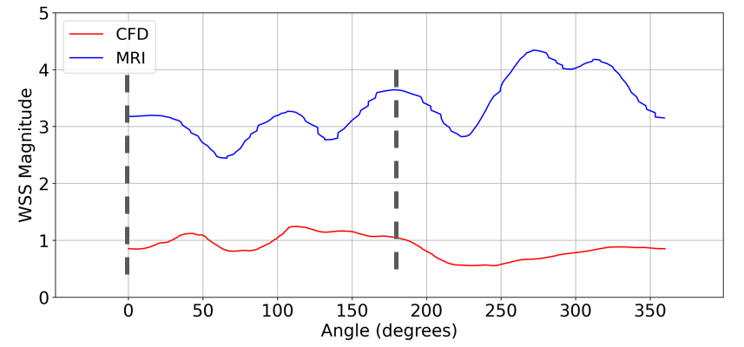}
        \caption{Circumferential WSS profile}
    \end{subfigure}
    \hfill
    \begin{subfigure}[t]{0.47\textwidth}
        \includegraphics[height=4.5cm]{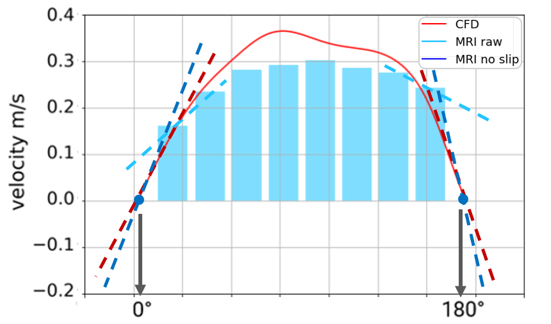}
        \caption{Velocity profiles along the wall-normal line}
    \end{subfigure}
    \caption{Case P04 (right), Slice~1. (a)~Circumferential WSS from CFD and from MRI with the no-slip constraint; dashed lines mark $0^{\circ}$ and $180^{\circ}$. (b)~Velocity profiles for raw MRI, MRI with no-slip condition, and CFD. In this case, imposing the no-slip condition steepened the near-wall gradient and led to a higher estimated WSS than CFD.}
    \label{sfig:velocity_gradient}
\end{figure}

In 4D flow MRI the vessel wall is not explicitly resolved, so measured velocities do not decay to zero at the lumen boundary. Enforcing zero wall velocity in the fit is physically correct but, with only two coarsely spaced sample points, can steepen the near-wall gradient. Velocity profiles were extracted in case P04 (right), Slice~1, along a line from the wall at $0^{\circ}$ across the cross-section to the opposite wall at $180^{\circ}$ (Figure~\ref{sfig:line_velocity}). Figure~\ref{sfig:velocity_gradient}a shows the WSS at $0^{\circ}$ and $180^{\circ}$ from CFD and from MRI with the no-slip constraint, and Figure~\ref{sfig:velocity_gradient}b compares the velocity profiles of CFD, raw MRI and MRI with the no-slip condition. The raw MRI profile is flat near the wall, giving a shallow gradient; in this example, the imposed no-slip constraint yielded a steeper fitted near-wall gradient and a higher estimated WSS than CFD.

% ─────────────────────────────────────────────────────────────────────────────
\section*{S5\quad Candidate Predictors and Multicollinearity Screening}
% ─────────────────────────────────────────────────────────────────────────────

\begin{figure}[H]
    \centering
    \includegraphics[width=0.85\linewidth]{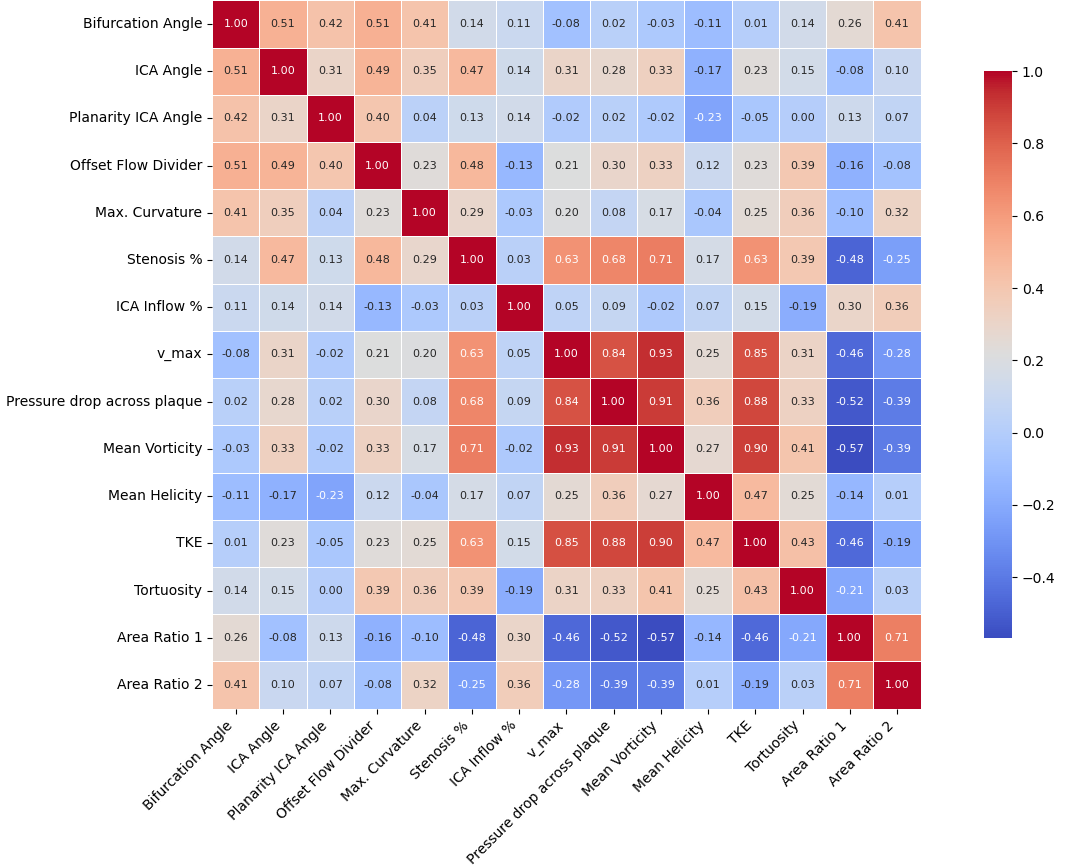}
    \caption{Pearson correlation matrix of the fifteen candidate predictors. Displayed names map to the manuscript terminology as follows: ``Area Ratio 1'' $= A_2/A_1$, ``Area Ratio 2'' $= A_3/A_1$, ``v\_max'' $=$ maximum velocity, ``Stenosis \%'' $=$ stenosis degree, ``ICA Inflow \%'' $=$ ICA inflow percentage, ``Tortuosity'' $=$ post-stenotic tortuosity, ``TKE'' $=$ turbulent kinetic energy.}
    \label{sfig:correlation_heatmap}
\end{figure}

Fifteen geometric and hemodynamic variables were extracted per artery (definitions in Table~\ref{stab:definitions} and in the companion study~\cite{mokhtari2025geometric}). Pearson correlations (Figure~\ref{sfig:correlation_heatmap}) and variance inflation factors (Table~\ref{stab:VIF_summary}) were computed for the arteries with complete data. Four hemodynamic variables (maximum velocity, pressure drop across the plaque, mean vorticity and turbulent kinetic energy) were strongly intercorrelated ($|r| > 0.7$), and three of them had VIF $> 10$; only pressure drop (VIF 8.85) was retained from this group. The two area ratios were also strongly correlated ($r = 0.71$) and only $A_2/A_1$ was retained. The resulting eleven-variable set was used in the regression models of Table~2 of the main manuscript.

\begin{table}[H]
\centering
\caption{Variance inflation factors of the fifteen candidate predictors before variable selection. Variables in bold (VIF $> 10$) were excluded; the area ratio $A_3/A_1$ was additionally excluded because of its correlation with $A_2/A_1$ ($r = 0.71$).}
\label{stab:VIF_summary}
\begin{tabular}{lc}
\toprule
\textbf{Variable} & \textbf{VIF} \\
\midrule
\textbf{Mean vorticity}                  & \textbf{28.71} \\
\textbf{Turbulent kinetic energy}        & \textbf{13.66} \\
\textbf{Maximum velocity}                & \textbf{10.93} \\
Pressure drop across plaque              & 8.85 \\
Area ratio $A_3/A_1$                     & 4.16 \\
Area ratio $A_2/A_1$                     & 3.29 \\
Bifurcation angle                        & 2.90 \\
Stenosis degree                          & 2.76 \\
Flow divider offset                      & 2.70 \\
Maximum curvature                        & 2.27 \\
ICA angle                                & 2.20 \\
Mean helicity                            & 2.14 \\
Post-stenotic tortuosity                 & 1.76 \\
ICA inflow percentage                    & 1.75 \\
Planarity of ICA angle                   & 1.61 \\
\bottomrule
\end{tabular}
\end{table}

\newpage
% ─────────────────────────────────────────────────────────────────────────────
\section*{S6\quad Parameter Definitions}
% ─────────────────────────────────────────────────────────────────────────────

Table~\ref{stab:definitions} defines 12 of the 15 candidate variables considered in the regression analyses; the remaining three (maximum velocity, mean vorticity and turbulent kinetic energy) and the full derivations are given in the companion study~\cite{mokhtari2025geometric}.

\begin{table}[H]
\centering
\small
\caption{Definitions of the candidate geometric and hemodynamic variables. Hemodynamic variables were extracted from the CFD solution at peak systole. The area ratio $A_3/A_1$, maximum velocity, mean vorticity and turbulent kinetic energy were excluded during multicollinearity screening; the latter three are defined in the companion study~\cite{mokhtari2025geometric}.}
\label{stab:definitions}
\begin{tabularx}{\linewidth}{lXl}
\toprule
\textbf{Variable} & \textbf{Definition} & \textbf{Unit} \\
\midrule
Stenosis degree & Relative area reduction $(1 - A_{\min}/A_{\mathrm{ref}})\times 100$ in the proximal ICA; $A_{\mathrm{ref}}$: reference lumen area distal to the stenosis & \% \\
Bifurcation angle & Angle between the ICA and ECA centerline tangents at the bifurcation point & $^{\circ}$ \\
ICA angle & Angle between the CCA and ICA centerline directions & $^{\circ}$ \\
Planarity of ICA angle & Out-of-plane angle of the ICA direction relative to the CCA-ECA plane & $^{\circ}$ \\
Flow divider offset & Distance of the flow divider from the CCA centerline axis, normalized by CCA diameter & dimensionless \\
Post-stenotic tortuosity & Centerline arc length divided by chord length in the ICA segment distal to the plaque & dimensionless \\
Maximum curvature & Maximum centerline curvature in the ICA & mm$^{-1}$ \\
Area ratio $A_2/A_1$ & Lumen area at Slice~2 divided by lumen area at Slice~1 & dimensionless \\
Area ratio $A_3/A_1$ & Lumen area at Slice~3 divided by lumen area at Slice~1 & dimensionless \\
Pressure drop across plaque & Difference in area-averaged pressure between cross-sections immediately proximal and distal to the plaque region & Pa \\
ICA inflow percentage & ICA flow rate as a percentage of CCA flow rate & \% \\
Mean helicity & Volume average of $\vec{v}\cdot(\nabla\times\vec{v})$ over the ICA segment containing the plaque & m/s$^2$ \\
\bottomrule
\end{tabularx}
\end{table}

\newpage
% ─────────────────────────────────────────────────────────────────────────────
\section*{S7\quad Correlation Results}
\begin{table}[htbp]
\centering
\caption{Spearman rank correlation coefficients between area-based stenosis degree and the signed CFD $-$ MRI difference (in m/s for velocity and Pa for WSS) at peak systole.}
\label{stab:spearman}
\begin{tabular}{llc}
\toprule
\textbf{Location} & \textbf{Metric} & $\rho$ \\
\midrule
Slice 2 & $v_\mathrm{mean}$ & 0.20 \\
Slice 2 & $v_\mathrm{max}$  & 0.36 \\
Slice 3 & $v_\mathrm{mean}$ & 0.28 \\
Slice 3 & $v_\mathrm{max}$  & 0.46 \\
\midrule
Slice 2 & Mean WSS & 0.24 \\
Slice 2 & Max WSS  & 0.16 \\
Slice 3 & Mean WSS & 0.29 \\
Slice 3 & Max WSS  & 0.45 \\
\midrule
Plaque surface & Mean WSS & 0.47 \\
Plaque surface & Max WSS  & 0.50 \\
\bottomrule
\end{tabular}
\end{table}

\printbibliography